\documentclass[showpacs,amssymb,prX]{revtex4}

\usepackage{amssymb,amsmath}
\usepackage{amsfonts}
\usepackage{color}
\usepackage{mathrsfs}

\def\ba{\begin{eqnarray}}
\def\ea{\end{eqnarray}}
\def\be{\begin{equation}}
\def\ee{\end{equation}}

\def\d{{\rm d}}

\begin{document}


\title{Uniqueness of the Fock quantization of a free scalar field
on $S^1$ with time dependent mass}
\author{Jer\'onimo Cortez}\email{jacq@fciencias.unam.mx}
\affiliation{Departamento de F\'\i sica,
Facultad de Ciencias, Universidad Nacional Aut\'onoma de M\'exico,
A. Postal 50-542, M\'exico D.F. 04510, Mexico.}
\author{Guillermo A. Mena Marug\'an}\email{mena@iem.cfmac.csic.es}
\affiliation{Instituto de Estructura de la Materia,
CSIC, Serrano 121, 28006 Madrid, Spain.}
\author{Rog\'erio Ser\^odio}\email{rserodio@mat.ubi.pt}
\affiliation{Departamento de Matem\'{a}tica, Universidade
da Beira Interior, R. Marqu\^es D'\'Avila e Bolama,
6201-001 Covilh\~a, Portugal.}
\author{Jos\'e M. Velhinho}\email{jvelhi@ubi.pt}
\affiliation{Departamento de F\'{\i}sica, Universidade
da Beira Interior, R. Marqu\^es D'\'Avila e Bolama,
6201-001 Covilh\~a, Portugal.}

\begin{abstract}

We analyze the quantum description of a free scalar
field on the circle in the presence of an explicitly
time dependent potential, also interpretable as a time
dependent mass. Classically, the field satisfies a
linear wave equation of the form
$\ddot{\xi}-\xi''+f(t)\xi=0$. We prove that the
representation of the canonical commutation relations
corresponding to the particular case of a massless
free field ($f=0$) provides a unitary implementation
of the dynamics for sufficiently general mass terms,
$f(t)$. Furthermore, this representation is uniquely
specified, among the class of representations
determined by $S^1$-invariant complex structures, as
the only one allowing a unitary dynamics. These
conclusions can be extended in fact to fields on the
two-sphere possessing axial symmetry. This generalizes
a uniqueness result previously obtained in the context
of the quantum field description of the Gowdy
cosmologies, in the case of linear polarization and
for any of the possible topologies of the spatial
sections.

\vskip 3mm \noindent
\end{abstract}
\pacs{04.62.+v, 03.70.+k, 04.60.Kz, 11.10.Kk}

\maketitle
\newpage
\renewcommand{\thefootnote}{\fnsymbol{footnote}}

\section{Introduction}
\label{intro}

In recent years, a Fock quantization allowing a
unitary dynamics of the linearly polarized Gowdy $T^3$
cosmological model \cite{gowdy} has been constructed
\cite{CM,ccm1,ccm2}. Moreover, this quantization has
been shown to be unique under very reasonable
conditions, namely, $S^1$-invariance and unitarity of
the dynamics for the metric field that describes the
local degrees of freedom of these spacetimes
\cite{ccmv,cmv}. These results are ground breaking,
since uniqueness in the quantization of cosmological
models is rare, and previous attempts on the
quantization of this Gowdy model \cite{pierri,torre}--
the simplest non-homogeneous cosmological model-- had
failed even in achieving a unitary implementation of
the dynamical evolution \cite{CM,cmv,torre,non-uni}.

Following the same methods and ideas, a Fock
quantization with unitary dynamics was later achieved
\cite{BVV1,BVV2} for the linearly polarized Gowdy
models with the other two allowed spatial topologies,
i.e., the $S^1\times S^2$ and $S^3$ models. Uniqueness
was proved again for these alternate topologies
\cite{cmv3} (see also the partial discussion in
\cite{BVV2}).

For any of the three considered topologies, the local
degrees of freedom of the model are parametrized by a
single scalar field, effectively living on $S^1$ in
the $T^3$ case, and on $S^2$ in the remaining cases.
Moreover, in all cases the dynamics is governed by
linear wave equations similar to those of the free
fields, though with a time dependent mass term.
Alternatively, the mass term can be regarded as a time
dependent potential.

On the one hand, it is clear that the compactness of
the effective space plays a role in allowing the
unitary implementation of the dynamics, and in the
uniqueness of the quantization \cite{ccmv,cmv} (see
also \cite{mtv} for a detailed account of the role
played in this respect by the long range behavior in
the non-compact case). On the other hand, it is not
known how the above results depend on the details of
the models, e.g., on the specific form of the mass
term appearing in the field equations. Since the
results are valid for the very different mass terms
appearing in the Gowdy $T^3$ model and in the
$S^1\times S^2$ and $S^3$ cases, as well as for a
constant mass term, of course, one may suspect that
the unitarity and uniqueness results may not be too
sensitive to the particular time dependence of the
mass. However, nothing in the works mentioned above
allows one to reach this conclusion, since methods
specially adapted to the specific mass term of each of
the models were used.

In the present work we will show that, for fields on
$S^1$ (as well as for axisymmetric fields on $S^2$,
see below), the commented results about the unitary
implementation of the dynamics and the uniqueness of
the representation are generic, i.e., they are valid
essentially as long as the dynamics is of the
specified type, regardless of the particular form of
the mass term - provided only that this term is given
by a sufficiently regular function.

From a more general perspective, it is important to
emphasize that the quantum representations in which
the dynamics is unitary coincide, in all cases of a
generic time dependent mass, with the representation
that is naturally associated with the free massless
field. So, the general belief that in quantum field
theory different dynamics require different
representations is not necessarily true in the compact
case, and therefore in some important cosmological
models. Somehow, the field with a time dependent mass
term possesses a dynamical behavior which -- in the
compact case -- is sufficiently close to the free
massless evolution, so that both dynamics are
implemented as unitary transformations in the same
representation.

This paper is organized as follows. First, we
summarize some aspects of the two-Killing vectors
reduction of General Relativity in Section 2, as a
motivation for the family of scalar fields that we are
going to consider. Then, we specify this family in
Section 3 and discuss its Fock quantization in the
representation that is naturally associated with the
free massless scalar field. In Section 4 we show that
this Fock quantization provides a unitary
implementation of our field dynamics, even if the
system contains a time dependent mass term. Section 5
proves that the considered Fock quantization is in
fact unique, inasmuch as it is the only Fock
representation that is invariant under the symmetry
group of the field equations and allows a unitary
quantum evolution. Finally, Section 6 presents our
conclusions.

\section{Motivation from the two-Killing vectors
reduction of General Relativity}

To motivate the interest of the systems that we will
study in this work, we will start by presenting a
brief summary on the reduction of General Relativity
to spacetimes that possess two commuting spacelike
Killing vectors. We consider only the case in which
the isometry group generated by these Killing vectors
is compact. Moreover, we restrict our discussion to
spacetimes for which each of these Killing vectors is
hypersurface orthogonal, a situation which is often
called the linearly polarized case.

In addition, we assume that the spacetime is globally
hyperbolic, so that it is possible to perform a 3+1
decomposition in sections of constant time $t$. Since
the isometry group generated by the Killing vectors is
Abelian (with non-null orbits), one can introduce
spatial coordinates $\{\theta,\phi,\rho\}$ such that
$\partial_{\phi}$ and $\partial_{\rho}$ are the
Killing vector fields, and the spacetime metric is
independent of $\phi$ and $\rho$ \cite{BCM}. As a
consequence of this independence, the integral $\int
\d \phi \d \rho=V_0$ (which is finite because the
isometry group is compact) appears as a global factor
in the gravitational action of General Relativity. We
absorb its numerical value in Newton's constant $G$ by
adopting units such that $8\pi G=V_0$.

On the other hand, one can fix the gauge freedom
associated with the momentum constraints (also called
diffeomorphism constraints) of the two directions
$\phi$ and $\rho$ by demanding the vanishing of the
components $h_{\theta\phi}$ and $h_{\theta\rho}$ of
the induced metric \cite{fix1,fix2}. This gauge fixing
\cite{fix2}, together with the assumption of
hypersurface orthogonality, implies that the metric
can be written globally in a diagonal form, except for
the presence of a $\theta$-component of the shift. The
reduced metric can be parametrized in the following
way \cite{CM}: \be \label{metric-1-red} d s^2 =
e^{\gamma - \psi}\left[-\tau^2
{N_{_{_{\!\!\!\!\!\!\sim}}\;}}^{2} d t^2 + (d \theta +
N^\theta d t)^2\right] + e^{- \psi}\tau^2(d \phi)^2 +
e^{\psi} (d \rho)^2. \ee

This reduced system still possesses two constraints:
the densitized Hamiltonian constraint,
$\widetilde{H}$, and the momentum constraint of the
$\theta$-direction, $H_{\theta}$. They take the
expressions:
\begin{eqnarray}\label{coh}
&&\widetilde{H}=\frac{1}{2} \left(\psi^{\prime}\tau
\right)^2+ \frac{1}{2}p_{\psi}^2
+\tau(2\tau^{\prime\prime}
-\tau^{\prime}\gamma^{\prime}-p_{\gamma}p_{\tau}).\\
\label{cohu} &&
H_{\theta}=-2p_{\gamma}^{\prime}+p_{\tau}
\tau^{\prime}+ p_{\gamma} \gamma^{\prime}+p_{\psi}
\psi^{\prime}.\end{eqnarray} In these formulas, the
$p$'s denote the momenta canonically conjugate to the
metric fields, and the prime stands for the spatial
derivative with respect to $\theta$. In principle, one
may introduce an additional gauge fixing to remove
these constraints and further reduce the system. The
particular details depend on those of the family of
spacetimes that one considers, such as the isometry
group and the spatial topology. Nonetheless, let us
admit for the moment that one can adopt (globally) a
gauge in which ${N_{_{_{\!\!\!\!\!\!\sim}}\;}}\tau=1$.
The generator of the time evolution is then the
integrated Hamiltonian constraint $\int \d \theta
(\widetilde{H}/\tau)$.

It is not difficult to check that the resulting
equation of motion for the field $\psi$ is \be
\ddot{\psi}+\frac{\dot{\tau}}{\tau}\dot{\psi}-
\frac{\tau^{\prime}}{\tau}\psi^{\prime}-\psi^{\prime\prime}
=0.\ee The dot denotes the derivative with respect to
$t$. Introducing the rescaling $\psi=(\xi
/\sqrt{\tau})y(\theta)$, where $y(\theta)$ is a fixed
function of the spatial coordinate $\theta$ only, the
above field equation translates into the following
equation for $\xi$: \be \label{xife}
\ddot{\xi}-\xi^{\prime\prime}-2\frac{y^{\prime}}{y}
\xi^{\prime}+f(t,\theta)\, \xi=0,\ee where \be
f(t,\theta)= \left(\frac{\dot{\tau}}{2\tau}\right)^2
-\frac{\ddot{\tau}}{2\tau}
-\left(\frac{\tau^{\prime}}{2\tau}\right)^2
+\frac{\tau^{\prime\prime}}{2\tau}
-\frac{y^{\prime\prime}}{y} .\ee

Obviously, Eq. (\ref{xife}) becomes a two-dimensional
wave equation with a spatially constant potential
$f=f(t)$ if $\tau$ is independent of $\theta$ and the
function $y$ is chosen equal to the unity. Explicitly,
\be \ddot{\xi}-\xi^{\prime\prime}+ f(t)\, \xi=0,\qquad
f(t)= \left(\frac{\dot{\tau}}{2\tau}\right)^2
-\frac{\ddot{\tau}}{2\tau}.\ee In addition, if $\tau$
admits an expression of the form
$\tau(t,\theta)=w(t)z(\theta)$, one gets a spatially
constant potential with the choice
$y(\theta)=\sqrt{z(\theta)}$. The field equation is
then \be\label{espwe}
\ddot{\xi}-\frac{1}{z(\theta)}\left[z(\theta)
\xi^{\prime}\right]^{\prime} + f(t)\, \xi=0,\qquad
f(t)= \left(\frac{\dot{w}}{2w}\right)^2
-\frac{\ddot{w}}{2w}.\ee At least locally, this
equation can be understood as the wave equation of a
$\phi$-independent field $\xi$ propagating in a
3-dimensional spacetime with the static metric \be
g_{ab}=-d t_a d t_b+d \theta_a d \theta_b+ z^2(\theta)
d \phi_a d \phi_b.\ee

The most interesting situation in which our previous
discussion finds a straightforward application is in
the case of the linearly polarized Gowdy cosmologies
\cite{gowdy}. These cosmological solutions are
(globally hyperbolic) spacetimes whose spatial
sections are compact and which possess two Killing
vector fields with the properties that we have
assumed. Actually, the spatial sections must be
homeomorphic to either the three-torus, $T^3$, the
three-sphere, $S^3$, or the three-handle, $S^2\times
S^1$. For these spacetimes, the gauge
${N_{_{_{\!\!\!\!\!\!\sim}}\;}}\tau=1$ is indeed
allowed \cite{CM,BVV1}. In the case of the topology of
the three-torus, this gauge is introduced by fixing
the freedom associated with the densitized Hamiltonian
constraint, what in turn is achieved by choosing the
metric function $\tau$ (essentially) as the time
coordinate, namely, $\tau=C t$ where $C$ is a constant
of motion \cite{CM,ccm1}. The corresponding field
$\xi$ is defined on the circle, while the potential of
the corresponding wave equation is given by the
function $1/(2t)^2$. The gauge fixing procedure can be
paralleled in the case of the other two topologies,
choosing $\tau(t,\theta)=C\sin{t}\sin{\theta}$, where
$C$ denotes again a constant of motion. In accordance
to our above discussion, the choice
$y=\sqrt{\sin{\theta}}$ leads then to a time dependent
potential, which turns out to be given by the function
$f(t)=(1+\csc^2{t})/4$. The corresponding field $\xi$
is defined on the sphere $S^2$, with the coordinate
$\theta$ being the zenith angle. This field is
axisymmetric, since its only spatial dependence is on
$\theta$. Finally, in these circumstances, the term
with spatial derivatives in the wave equation
(\ref{espwe}) can be interpreted just as the Laplacian
on the two-sphere acting on the axisymmetric field
$\xi$.

\section{Fock quantization of the model}
\label{model}

\subsection{The classical model}

After the above motivation, we can now specify the
models on which we will concentrate our discussion.
The system which we want to study is a scalar field
{$\xi(t,\theta)$} propagating in a $(1+1)$-spacetime
with the topology of $\mathbb{I} \times S^1$ and
provided with the static metric
$g_{ab}=-dt_{a}dt_{b}+d\theta_{a} d\theta_{b}$. The
time domain $\mathbb{I}$ is an interval of the real
line and, in most of the practical situations, will be
taken to coincide either with $\mathbb{R}^{+}$ or with
$\mathbb{R}$. In addition, the field is subject to a
time-dependent potential of the form
$V(\xi)=f(t)\xi^{2}/2$, where $f(t)$ is a
{sufficiently regular} function on the interval
$\mathbb{I}$. As we have already pointed out, this
potential can be thought of as a time-dependent mass
term. Later in this work we will comment on the
extension of our analysis to the case of an
axisymmetric field on $S^2$, instead of a field on the
circle.

In the canonical approach, in principle, the system
can be described by the action \be \label{hamil}
S(t_i,t_f)=\int_{t_{i}}^{t_{f}}\d t\,
\left[\left(\oint \d \theta\,
P\dot{\varphi}\right)-H\right],\quad
H=\frac{1}{2}\oint \d \theta \,
\left[P^{2}+(\varphi^{\prime})^{2}+f\varphi^{2}\right],
\ee where $H$ is the Hamiltonian and $\varphi$ and $P$
are, respectively, the configuration and momentum of
the field $\xi$.

The canonical phase space is the space of Cauchy data
$\{(\varphi,P)\}=\{(\xi_{|t_0}, {\dot \xi}_{|t_0})\}$
at some fixed time $t_0$. The corresponding nonzero
Poisson brackets are
$\{\varphi(\theta),P(\theta')\}=\delta(\theta-\theta')$,
where $\delta(\theta)$ is the Dirac delta on $S^1$.
Varying the action (\ref{hamil}) with respect to
$\varphi$ and $P$, we arrive to the field equations
\begin{equation}
\label{2}
 \dot \varphi=P,\qquad\
\dot P=\varphi^{\prime\prime} -f(t)\varphi,
\end{equation}
so that $\xi$  {satisfies} the linear wave equation
\begin{equation}
\label{1} \ddot{\xi}-\xi^{\prime\prime}+f(t)\xi=0.
\end{equation}
Alternatively to the space of Cauchy data
$\Gamma=\{(\varphi, P)\}$, the phase space can be
described as the space $S=\{\xi\}$ of solutions to Eq.
(\ref{1}). Both $\Gamma$ and $S$ are symplectic linear
spaces, with the respective symplectic structures
(independent of the choice of time section)
\begin{equation}
\sigma[(\varphi_{1},P_{1}),(\varphi_{2},P_{2})] =\oint
\d \theta \, \left(\varphi_{2}P_{1}
-\varphi_{1}P_{2}\right),\end{equation} and
\begin{equation} \Omega(\xi_{1},\xi_{2})=\oint
\d \theta \, \left(\xi_{2}\partial_{t}\xi_{1}
-\xi_{1}\partial_{t}\xi_{2}\right).\end{equation}

Since the Hamiltonian (\ref{hamil}) does not depend on
the spatial variable $\theta$, the field equations are
invariant under $S^1$-translations: \be
\label{s1-trans} T_{\alpha}:\theta\mapsto
\theta+\alpha \qquad \forall \alpha \in S^{1}. \ee So,
the translations $T_{\alpha}$ form a group of
symmetries of the dynamics
$^{\footnotemark[1]}$\footnotetext[1]{Moreover, in
situations such as the case of the linearly polarized
Gowdy $T^3$ cosmologies, these symmetries are in fact
gauge transformations of the reduced system obtained
after an almost complete gauge fixing \cite{ccmv}.}.

Given the periodicity in the spatial coordinate
$\theta$, one can equivalently use the Fourier
coefficients of $\varphi$ and $P$ as coordinates of
our phase space. Let us be more explicit: employing
the Fourier series expansion of $\varphi$ and $P$
\begin{eqnarray}
\label{3} \varphi&=&{\frac{q_0}{\sqrt{2\pi}}}+
\sum_{n>0} \left(q_n
{\frac{\cos(n\theta)}{\sqrt{\pi}}} +
x_n {\frac{\sin(n\theta)}{\sqrt{\pi}}}\right),\\
\label{4} P&=&{\frac{p_0}{\sqrt{2\pi}}}+ \sum_{n>0}
\left(p_n {\frac{\cos(n\theta)}{\sqrt{\pi}}}+ y_n
{\frac{\sin(n\theta)}{\sqrt{\pi}}}\right),
\end{eqnarray}
we see that the space of Cauchy data $\{(\varphi ,
P)\}$ is in a one-to-one correspondence with the space
of real Fourier coefficients $\{(q_n , x_n , p_n ,
y_n);\, n\in \mathbb{N}^+\}\bigcup \{(q_0,p_0)\}$.
Besides, from the basic Poisson bracket between
$\varphi$ and $P$, one can check that
$\{q_0,p_{0}\}=1$ and
$\{q_n,p_{n'}\}=\{x_n,y_{n'}\}=\delta_{nn'}$, the rest
of brackets being equal to zero. Thus, the canonical
phase space can be coordinatized either by
$\{(\varphi,P)\}$ or by the set of canonical pairs
$\{(q_n,p_{n}),(x_n,y_{n});\, n\in
\mathbb{N}^+\}\bigcup \{(q_0,p_0)\}$.

By substituting the expansions (\ref{3}) and (\ref{4})
in Eq. (\ref{2}) we obtain
\begin{equation}
\label{qn-pn-eq}
\dot q_n=p_n,\qquad \dot p_n=-\left(n^2+f\right)q_n,
\end{equation}
and therefore \be \label{q-eq} \ddot
q_n+\left(n^2+f\right)q_n=0. \ee  These formulas are
also valid for $q_0$ and $p_0$, just by letting $n$
vanish. In addition, by replacing $q_{n}$ with $x_{n}$
and $p_{n}$ with $y_{n}$ (with $n>0$), one obtains the
equations of motion corresponding to the pair
$(x_n,y_n)$.

Let us introduce now the complex phase space variables
\begin{equation}
\label{basic-var}
a_n={\frac{1}{\sqrt{2n}}}(nq_n+ip_n),\qquad \tilde
a_n={\frac{1}{\sqrt{2n}}}(nx_n+iy_n), \qquad
n\in\mathbb{N}^+.
\end{equation}
These are just the usual annihilation-like variables
for a system of harmonic oscillators with frequencies
equal to $n$. Of course, that system corresponds to
the particular case $f=0$ in Eq. (\ref{1}), i.e., to
the free massless field case. We can use the above set
of complex variables, together with the associated
creation-like variables obtained by complex
conjugation, in order to coordinatize the
inhomogeneous sector ($n\neq 0$) of the canonical
phase space. Besides, from now on, we ignore the zero
mode ($n=0$) in our analysis, since its dynamics is
decoupled from that of the inhomogeneous sector, the
mode can be quantized by standard methods, and it
plays no role in the subsequent discussion of
unitarity and uniqueness of the quantum
representation.

For the nonzero modes, the dynamics is dictated by the
inhomogeneous part $\bar{H}$ of the Hamiltonian
(\ref{hamil}), which in terms of our new set of
variables adopts the expression: \be
\label{inhomo-hamil}
\bar{H}=\sum_{n>0}\left[\left(n+\frac{f}{2n}\right)\,
(a^{*}_{n} a_{n} +
\tilde{a}^{*}_{n}\tilde{a}_{n})+\frac{f}{4n}
\left(a_{n}a_{n}+\tilde{a}_{n}\tilde{a}_{n}+
a^{*}_{n}a^{*}_{n}+\tilde{a}^{*}_{n}
\tilde{a}^{*}_{n}\right)\right]. \ee The symbol $*$
denotes complex conjugation.

The finite transformations generated by the
Hamiltonian $\bar{H}$ are linear symplectic
transformations which can be decomposed in $2\times 2$
blocks, one for each fixed pair
${\cal{A}}_{n}=(a_n,a^{*}_n)$ and
$\tilde{{\cal{A}}}_{n}=(\tilde{a}_n,\tilde{a}^{*}_n)$.
Furthermore, the blocks for these two pairs (with the
same mode number $n$) coincide. Thus, the classical
evolution of the annihilation and creation-like
variables from time $t_0$ to time $t$ is totally
determined by a sequence of $2\times 2$ matrices
${\cal U}_n(t,t_0)$, with $n\in\mathbb{N}^+$, of the
form:
\begin{eqnarray}
\left( \begin{array}{c} a_n (t)\\ a_n^*(t)
\end{array}\right)
= {\cal U}_n(t,t_0)\left( \begin{array}{c} a_n (t_0)\\
a_n^* (t_0)\end{array} \right), \qquad  \left(
\begin{array}{c} \tilde{a}_n (t)\\ \tilde{a}_n^*(t)
\end{array}\right)
= {\cal U}_n(t,t_0)\left( \begin{array}{c}
\tilde{a}_n(t_0)
\\ \tilde{a}_n^*(t_0) \end{array} \right),
\end{eqnarray}
\begin{eqnarray}
\label{bogo-transf}
{\cal U}_n(t,t_0)=\left( \begin{array}{cc} \alpha_n(t,t_0)
& \beta_n(t,t_0) \\ \beta_n^*(t,t_0) &
\alpha_n^*(t,t_0) \end{array} \right),
\end{eqnarray}
where $\alpha_n(t,t_0)$ and $\beta_n(t,t_0)$ are
Bogoliubov coefficients which depend on the specific
function $f(t)$ of the system.

At this stage of the discussion, a couple of remarks
about the extension of our analysis are in order.

\begin{enumerate}
\item{In the field models of the considered type which arise
from symmetry reductions of General Relativity to
cosmological scenarios, the time interval $\mathbb{I}$
is contained, typically, in the positive semiaxis
$\mathbb{R}^{+}$. The origin $t=0$ corresponds to a
big bang singularity. This singularity generically
implies that the function $f$ is no longer well
behaved at that point. In cases like the Gowdy $T^3$
cosmologies, the time domain is unlimited in the
evolution of the system apart from this singularity,
and thus $\mathbb{I}$ coincides with $\mathbb{R}^{+}$.
On the other hand, the function $f$ may have more than
one singularity, and in this case the time domain is
further restricted to a bounded interval, as it
happens to be the case for the Gowdy $S^1\times S^2$
and $S^3$ cosmological models, discussed below when
one allows the compact spatial (topological) manifold
to differ from the circle. Obviously, one can consider
more general settings than these models inspired in
cosmology. In particular, for smooth functions $f$ one
can extend its domain of definition $\mathbb{I}$ to
the whole real line, and the study applies then to
scalar fields in $\mathbb{R}\times S^1$.}

\item{As we have anticipated, our discussion can be
extended to axisymmetric fields on $S^2$. With this
aim, let us start by considering the action of a free
scalar field on $\mathbb{I}\times S^2$ in the presence
of a time dependent potential. This action is of the
form (\ref{hamil}) with the spatial integration
performed over $S^2$ instead of over the circle, and
with the quadratic term in spatial derivatives,
$(\varphi^{\prime})^2$, of the Hamiltonian replaced
with $\eta^{ij}\partial_i\varphi
\partial_j\varphi$, where the indices $i$ and $j$ denote
spatial indices on the sphere, and $\eta_{ij}$ is the
round metric on $S^2$, namely, $\eta_{ij}=d\theta_i
d\theta_j+\sin{\theta^2} d\phi_id\phi_j$. The
resulting field equation is similar to Eq. (\ref{1}),
but with the second spatial derivative replaced with
$\Delta \xi$, $\Delta$ being the Laplace-Beltrami
operator on the two-sphere. The expansion of the
Cauchy initial data is now performed in terms of
spherical harmonics, $Y_{lm}(\theta,\phi)$, which are
eigenfunctions of the operator $\Delta$ with
eigenvalue equal to $l(l+1)$. The requirement of
axisymmetry restricts the harmonics in this expansion
to the set $\{ Y_{l0}, l\in \mathbb{N}\}$, the only
spherical harmonics which are independent of $\phi$.
The coefficients of $\xi$ in this expansion in
harmonics satisfy an equation identical to
(\ref{q-eq}), except for the substitution of $n$ by
$l+\frac{1}{2}$ and the redefinition of the function
$f(t)$, whose role is played now by
$\bar{f}(t):=f(t)-1/4$. From this point on, the
discussion is completely parallel to that presented
for the field on the circle, with the only caveat that
the mode numbers $n$ correspond now to positive
half-integers, $l+\frac{1}{2}$, rather than to
positive integers, a fact which, nonetheless, does not
affect the computations nor the rationale of our
analysis.}
\end{enumerate}

\subsection{Complex structure}

Let us start our discussion about the quantization
process by considering the choice of complex structure
for the system, which is the mathematical structure
that encodes all the ambiguity which is physically
relevant in the Fock quantization. Recall that a
complex structure $J$ is a symplectic transformation
in $S$ (or in $\Gamma$), compatible with the
symplectic structure [in the sense that their
combination $\Omega(J\cdot,\cdot)$ provides a positive
definite bilinear map], and such that its square
equals minus the identity, $J^2=-1$.

The time evolution can be viewed as a map that relates
copies of $\Gamma$ at different times, e.g.,
$\{({\cal{A}}_{n}(t_{0}),
\tilde{{\cal{A}}}_{n}(t_{0}));\, n\in\mathbb{N}^+\}$
at $t_{0}$ with $\{({\cal{A}}_{n}(t),
\tilde{{\cal{A}}}_{n}(t));\,n\in\mathbb{N}^+\}$ at
$t$. For definiteness, we choose once and for all an
initial reference time $t_{0}$. Associated with
$\{({\cal{A}}_{n}(t_{0}),
\tilde{{\cal{A}}}_{n}(t_{0}));\, n\in\mathbb{N}^+\}$
[which we will denote by $\{({\cal{A}}_{n}(t_{0}),
\tilde{{\cal{A}}}_{n}(t_{0}))\}$ in the following, to
simplify the notation], there is a natural field
decomposition $\xi=\xi^{+}+\xi^{-}$, where \be
\xi^{+}(t,\theta)=\sum_{n>0}\frac{1}{\sqrt{2\pi
n}}M_{n}(t)\left[\cos(n\theta)a_{n}(t_{0})+
\sin(n\theta)\tilde{a}_{n}(t_{0})\right], \ee
$\xi^{-}$ is the complex conjugate of $\xi^{+}$, and
$M_{n}(t)=\alpha_{n}(t,t_{0})+\beta^{*}_{n}
(t,t_{0})$. Expressing the cosine and sine functions
in terms of exponentials, we can rewrite $\xi^{+}$ as
\be \label{positive} \xi^{+}(t,\theta)=\sum_{n\neq
0}\frac{1}{\sqrt{4\pi |n|}}M_{|n|}(t)\,
e^{in\theta}b_{n}, \ee where
$b_n:=\frac{1}{\sqrt{2}}[a_n (t_{0})-i\tilde a_n
(t_{0})]$ and $b_{-n}:=\frac{1}{\sqrt{2}}[a_n
(t_{0})+i\tilde a_n (t_{0})]$. The explicit
decomposition of the solutions in complex conjugate
pairs defines the $\Omega$-compatible complex
structure $J_{0}$ on $S$:
\begin{equation}J_{0}(M_{|n|}(t)e^{in\theta})=
iM_{|n|}(t)e^{in\theta}, \qquad
J_{0}(M^{*}_{|n|}(t)e^{-in\theta})=
-iM^{*}_{|n|}(t)e^{-in\theta}.\end{equation} In the
$\{({\cal{A}}_{n}(t_{0}),
\tilde{{\cal{A}}}_{n}(t_{0}))\}$ basis, the complex
structure $J_{0}$ is given by a block diagonal matrix,
with $4\times 4$ blocks
$(J_{0})_n={\rm{diag}}(i,-i,i,-i)$ for each value of
$n$.

Since the reference time $t_{0}$ has been chosen
arbitrarily in $\mathbb{I}$, we can reproduce our
analysis for any other time value in this interval,
let's say $t=T$. In that case, the field $\xi(t,
\theta)$ will be decomposed again in ``positive'' and
``negative'' frequency solutions analog to $\xi^+$ and
$\xi^-$ [see Eq. (\ref{positive})], but now in terms
of the coefficients $b_{n}(T)$ and the modes
$M^{T}_{|n|}(t)$ which, in turn, are obtained by
replacing $t_{0}$ with $T$ in the expressions given
above for $b_n$, $b_{-n}$, and $M_{|n|}(t)$. Thus, for
each copy $\{({\cal{A}}_{n}(T),
\tilde{{\cal{A}}}_{n}(T))\}$ of $\Gamma$ [which,
alternatively, can be coordinatized by
$\{{\cal{B}}_{n}(T)\}:=\{(b_n(T),b^{*}_{-n}(T),
b_{-n}(T),b^{*}_n(T))\}$], we obtain a natural field
decomposition, and hence a natural complex structure
$J_{T}$ (note that we have called $J_{t_{0}}=J_0$ to
simplify the notation). In this way, we arrive at a
uniparametric family of solution spaces of positive
[negative] frequency,
$S_{T}^{+}:=\{\xi^{+}=(\xi-iJ_{T}\xi)/2\}$
[$S_{T}^{-}=(S_{T}^{+})^*$], which is induced by the
evolution.

Note also that, in the alternative $\{{\cal{B}}_{n}\}$
description of $\Gamma$, the time evolution is
dictated precisely by the sequence of $2\times 2$
matrices ${\cal U}_n(t,t_0)$, though now acting on the
pairs $(b_n,b^{*}_{-n})$ and $(b_{-n},b^{*}_n)$. In
the $\{{\cal{B}}_{n}\}$ basis, the complex structure
$J_{0}$ is also given by a block diagonal matrix, with
the $4\times 4$ blocks
$(J_{0})_n={\rm{diag}}(i,-i,i,-i)$.

Finally, it is worth recalling that, in the free
massless field case, one has
$\alpha_n(t,t_0)=e^{-in(t-t_0)}$ and
$\beta_n(t,t_0)=0$ (with $n>0$), so that the positive
mode solutions $M_{|n|}(t)e^{in\theta}$ associated
with the complex structure $J_{0}$ in the free case
are simply the usual subfamily of plane waves
$e^{-i|n|(t-t_0)+in\theta}$.

\subsection{Quantum representation}

Starting with $(S,J_0)$  one can construct in a
standard way the Hilbert space of the quantum theory.
The first step is to complete the space of positive
frequency solutions specified by $J_0$ with respect to
the norm $\sqrt{\langle \xi^{+} , \xi^{+}\rangle}$,
where $\langle \xi^{+}_{1} ,
\xi^{+}_{2}\rangle:=-i\Omega[(\xi^{+}_{1})^* ,
\xi^{+}_{2}]$. The result is the so-called
one-particle Hilbert space, ${\mathscr{H}}$. Next,
{the Hilbert space is obtained} by considering the
symmetrized tensor product of $n$ copies of
${\mathscr{H}}$, one for each $n\in \mathbb{N}$, and
{collecting} the resulting spaces via the direct sum
operation. In short, the Hilbert space {is  the} Fock
space associated with ${\mathscr{H}}$:
$${\mathscr{F}}({\mathscr{H}})=\oplus_{n=0}^{\infty}
\left(\otimes_{(s)}^{n}{\mathscr{H}}\right).$$ In this
prescription, the field operator $\hat{\xi}$ is
written in terms of the annihilation and creation
operators corresponding to the positive and negative
parts defined by the complex structure $J_0$, namely,
\be \hat{\xi}(t;\theta)=\sum_{n\neq
0}\frac{1}{\sqrt{4\pi |n|}}M_{|n|}(t)\,
e^{in\theta}\hat{b}_{n}+{\rm{h.c.}}, \ee where
``h.c.'' stands for ``Hermitian conjugate''. One can
also rewrite the field operator in terms of the
annihilation $\{\hat{a}_{n},\hat{\tilde{a}}_{n}\}$ and
creation
$\{\hat{a}^{\dag}_{n},\hat{\tilde{a}}^{\dag}_{n}\}$
operators associated with the positive and negative
frequency solutions corresponding to excitations of
the ``oscillators'' $q_{n}$ and $x_{n}$: \be
\hat{\xi}(t;\theta)=\sum_{n>0}\frac{1}{\sqrt{2\pi
n}}M_{n}(t)\left[\cos(n\theta)\hat{a}_{n}(t_{0})+
\sin(n\theta)\hat{\tilde{a}}_{n}(t_{0})\right] +
{\rm{h.c.}} \ee In the Heisenberg picture, time
evolution is, in principle, provided by the Bogoliubov
transformation (\ref{bogo-transf}), what means that
one can define operators $\hat{a}_{n}(t),
\hat{a}^{\dag}_{n}(t)$ at time $t$, related with the
operators $\hat{a}_{n}(t_0), \hat{a}^{\dag}_{n}(t_0)$
at time $t_0$ according to
\begin{eqnarray}
\left( \begin{array}{c} \hat{a}_n(t) \\
\hat{a}^{\dag}_n (t)\end{array}\right) = \left(
\begin{array}{cc} \alpha_n(t,t_0) & \beta_n(t,t_0) \\
\beta_n^*(t,t_0) &  \alpha_n^*(t,t_0) \end{array}
\right) \, \left( \begin{array}{c}  \hat{a}_n(t_0) \\
\hat{a}^{\dag}_n (t_0)  \end{array}\right),
\end{eqnarray}
and a completely similar expression for
$(\hat{\tilde{a}}_{n}, \hat{\tilde{a}}^{\dag}_{n})$.

A key question is to elucidate whether the above
transformations correspond to unitary transformations
in ${\mathscr{F}}({\mathscr{H}})$, i.e., whether or
not the dynamics is implementable in a unitary way on
the Fock representation determined by $J_0$ (in the
following, we will call it the $J_0$-Fock
representation). Let us recall that a symplectic
transformation $R$ can be unitarily implemented on a
Fock representation, constructed from a complex
structure $J$, if and only if $(R+JRJ)$ is an operator
of the Hilbert-Schmidt type on the corresponding
one-particle Hilbert space \cite{hr, sh}.
Equivalently, $R$ is implementable as a unitary
transformation if and only if the representations
defined by $J$ and $RJR^{-1}$ are unitarily
equivalent, i.e., if and only if $(J- RJR^{-1})$ is
Hilbert-Schmidt.

In the case of the family of symplectic
transformations ${\cal U}(t,t_{0})$ defined by the
classical dynamics [and specified by the matrices
${\cal U}_n(t,t_{0})$], the Hilbert-Schmidt condition
for a unitary implementation in the $J_0$-Fock
representation becomes a square summability condition
on the coefficients $\beta_n$, namely,
$\sum_{n=1}^{\infty}|\beta_{n}(t,t_{0})|^{2}<\infty$
$\forall t\in \mathbb{I}$, given a fixed reference
time $t_0$. Before proving that this condition is
indeed satisfied, let us conclude the subsection with
some additional comments.

\begin{enumerate}
\item{In order to construct the Fock representation,
we could have considered the space $S^{+}_{T}$,
determined by the complex structure $J_T$, rather than
the space of positive frequency solutions specified by
$J_0$. Since the time $T$ can take any value in
$\mathbb{I}$, we would have obtained in this way a
uniparametric family of $J_T$-Fock representations.
Clearly, the $J_0$-Fock representation belongs to this
family and corresponds to $T=t_0$. Note that unitary
implementability of the dynamics on the $J_0$-Fock
representation (and actually on any representation
within the family) amounts to the unitary equivalence
of all the $J_T$-Fock representations.}

\item{
By considering the counterpart of $J_0$ on the
canonical phase space $\Gamma$ (rather than on $S$),
one can construct the functional representation which
is unitarily equivalent to the $J_0$-Fock description
(see \cite{ccmv2} for a detailed treatment in complex
variables and \cite{ccq2} for the GNS relationship
between Schr\"{o}dinger and Fock representations). The
result is a Schr\"{o}dinger representation of the
canonical commutation relations on the Hilbert space
${\cal H}=L^2({\cal Q},\mu)$ of square integrable
functions on the infinite dimensional linear space
${\cal Q}=\{(q_n,x_n);\quad n\in\mathbb{N}^+\}\cong
(\mathbb{R}^2)^{\mathbb{N}^+}$, with respect to the
Gaussian measure
\begin{equation}
\d \mu=\prod_{n>0}\left(e^{-n(q_n^2+x_n^2)}\,
\,\frac{n}{\pi}\,\d q_n \d x_n\right).
\end{equation}
The basic operators of configuration ($\hat{q}_n$ and
$\hat{x}_n$) and momentum ($\hat{p}_n$  and
$\hat{y}_n$) act as multiplicative and derivative
operators, respectively:
\begin{eqnarray}
\hat q_n \Psi &= q_n\Psi ,\qquad\qquad\qquad &\hat x_n
\Psi=
x_n\Psi , \\
\hat p_n\Psi &= -i {\frac{\partial}{\partial q_n}}\Psi
+i n q_n\Psi, \qquad &\hat y_n\Psi= -i
{\frac{\partial}{\partial x_n}}\Psi+i n x_n\Psi.
\end{eqnarray} Here,
$\Psi\in {\cal H}$ is an arbitrary
``wave function''.

If one employs relation (\ref{basic-var}) to introduce
operators $(\hat{a}_n , \hat{\tilde{a}}_n)$ and
$(\hat{a}^{\dag}_n , \hat{\tilde{a}}^{\dag}_n)$, it is
easy to check that these provide the annihilation and
creation operators of the representation. Thus, the
constructed representation is just the one which is
naturally associated with the free massless field
case. A different way to see this fact is by computing
the counterpart of the complex structure $J_0$ on the
canonical phase space, $j_0$, which in terms of the
variables $(\varphi,P)$ takes the familiar form
\begin{eqnarray}
\label{cano-cs}
j_0= \left( \begin{array}{cc} 0 & -(-\Delta)^{-1/2}
\\(-\Delta)^{1/2} &  0\end{array}\right).
\end{eqnarray}
Obviously, the massless free field dynamics is
implemented as a unitary transformation in this
representation; actually, the corresponding
coefficients $\beta_n$ vanish identically, since the
complex structure is invariant under the free field
dynamics. Nevertheless, in the next section we will
prove a nontrivial result, namely, that the dynamics
of the field $\xi$ that we are studying admits also a
unitary implementation in the considered
representation tailored to the free massless field.}
\end{enumerate}

\section{Unitary dynamics}
\label{sec:unit}

In this section, we want to address the question of
whether the sequences $\{\beta_{n}\}$ are square
summable or not. Thus, we will be interested in the
large $n$ limit of the coefficients $\beta_n$, and
therefore in the behavior of the equations of motion
(\ref{q-eq}) for large $n$.

Let us start by writing the general solution to those
equations of motion in the form \be \label{gen}
q_n(t)=A_n\exp[n\Theta_n(t)]+
A^{*}_n\exp[n\Theta^{*}_n(t)], \ee where, for each
$n$, $A_n$ is a complex constant and $\Theta_n$ is a
particular complex solution of the characteristic
equation \be \label{car} n\ddot \Theta_n+n^2{\dot
\Theta_n}^2+n^2+f=0, \ee arising from Eqs.
(\ref{q-eq}) and (\ref{gen}).

A simple calculation shows the relation between
$\Theta_n(t)$ and the modes $M_n(t)$ associated with
the complex structure $J_0$: \be \label{rel-theta-m}
M_n(t)=-\exp\{n[\Theta_n(t)-\Theta_n(t_0)]\}
\frac{1-i\dot{\Theta}_n^{*}(t_0)}{2{\rm{Im}}
\dot{\Theta}_n(t_0)}+\exp\{n[\Theta_n^{*}(t)-
\Theta_n^{*}(t_0)]\}\frac{1-i\dot{\Theta}_n^{*}
(t_0)}{2{\rm{Im}}\dot{\Theta}_n (t_0)}. \ee

It is worth pointing out that Eq. (\ref{car}) involves
only the function $\dot \Theta_n$ and its derivative.
Actually, it is just a first-order differential
equation of the Riccati type for $\dot \Theta_n$.
Hence, the functions $\Theta_n$ are determined only up
to additive constants. We use this freedom to set
$\Theta_n(t_0)=0$. Let us consider now the freedom in
choosing a particular solution to Eq. (\ref{car}) for
each $n\in\mathbb{N}^+$. By computing the relation
between the initial data $(q_n(t_0), p_n(t_0))$, on
the one hand, and $\dot{\Theta}_n(t_0)$ and the
complex arbitrary constants $A_n$ appearing in Eq.
(\ref{gen}), on the other hand, one can check that it
is possible to reach any value of the initial data
while setting $\dot{\Theta}_n(t_0)=-i$. This condition
fixes then the solution to Eq. (\ref{car}). The choice
is motivated by our knowledge of the free massless
scalar field, case in which $\dot{\Theta}_n=-i$ is
satisfied not only initially, but at all times.
Substituting the resulting relation between
$(q_n(t_0), p_n(t_0))$ and $A_n$ in Eq. (\ref{gen}),
it is easy to obtain the evolution matrices in terms
of the original variables $(q_n,p_n)$. Changing from
those variables to the annihilation and creation-like
variables $(a_{n},a_n^*)$, one can deduce the
expression of the Bogoliubov coefficients
$\alpha_{n}(t,t_0)$ and $\beta_{n}(t,t_0)$ as
functions of the real  and imaginary parts of
$\Theta_n(t)$, which we call $r_n(t)$ and $s_n(t)$,
respectively: \begin{eqnarray} \label{alpha}
\alpha_n(t,t_0) & = & \frac{1}{2}e^{nr_n(t)}
e^{ins_n(t)}\left[1+i\,\dot
r_n(t)-\dot s_n(t) \right] ,\\
\label{beta} \beta_n(t,t_0) & = &
\frac{1}{2}e^{nr_n(t)} e^{-ins_n(t)}\left[1+i\,\dot
r_n(t)+\dot s_n(t) \right] .
\end{eqnarray}
In the equations of motion (\ref{q-eq}), the $n^2$
term dominates over the mass term $f(t)$ in the limit
of large $n$ modes, and  we thus expect that the
solutions $q_n(t)$ converge to those corresponding to
the massless case for the same initial conditions, at
least for sufficiently regular functions $f(t)$ on
$\mathbb{I}$. Hence, the exponential $e^{nr_n(t)}$
tends to 1, and thus $\beta_n(t,t_0)$ is square
summable if and only if so is
$2e^{-nr_n(t)}\beta_n(t,t_0)$. Therefore, in the
following we will focus our analysis on the behavior
of $\dot{r}_n$ and $\dot{s}_n$ for large $n$.

Let us write the functions $\dot{\Theta}_n$ in the
form \be \label{newze1}
\dot{\Theta}_n=-i+\frac{W_n}{n}. \ee The initial
condition on $\dot{\Theta}_n$ translates then into the
vanishing of $W_n$ at $t_0$. Besides, from Eq.
(\ref{car}), it follows that the functions $W_n$
satisfy the first-order differential equations \be
\label{car-w} \dot{W}_{n}=2in W_{n}- W^{2}_{n}-f, \ee
also of the Riccati type. We want to show now that, in
the large $n$-limit, the desired solutions to Eq.
(\ref{car-w}) admit ``ultraviolet modes'' of order
$1/n$. Thus, in particular, the sequences $W_n(t)$
tend to zero, and the sequences $W_n(t)/n$ are square
summable $\forall t$. The argument is the following.
In the asymptotic limit of large $n$, the quadratic
term $W^{2}_{n}$ in Eq. (\ref{car-w}) is expected to
be dominated by the linear term in $W_{n}$, whose
coefficient is proportional to $n$ and therefore grows
in the asymptotic regime under consideration. We will
hence neglect that quadratic term, show that the
resulting linear equation admits solutions $\bar{W}_n$
of order $1/n$, and check that, in the asymptotic
limit $n\rightarrow \infty$, the contribution of the
quadratic term for such solutions is in fact
negligible in our original differential equation.

Thus, let us consider the linear equation obtained
from Eq. (\ref{car-w}) after removing the quadratic
term $W^{2}_{n}$: \be \label{w-lin}
\dot{\bar{W}}_{n}=2in \bar{W}_{n}-f. \ee The solution
to Eq. (\ref{w-lin}) satisfying the initial condition
$W_n(t_0)=0$ is given by \be
\bar{W}_{n}(t)=-\exp(2int)\,\int_{t_0}^t \d \bar{t}\,
f(\bar{t})\exp(-2in\bar{t}). \label{lin-sol} \ee A
simple integration by parts leads then to \be
\label{newze2}
\bar{W}_{n}(t)=-\frac{if(t)}{2n}+\frac{if(t_0)\,
e^{2in(t-t_0)}}{2n} -\frac{\exp(2int)}
{2in}\,\int_{t_0}^t \d \bar{t}\,
\dot{f}(\bar{t})\exp(-2in\bar{t}), \ee and one can
easily check that the absolute value of the last
term is bounded by $\frac{1}{2n}\int_{t_0}^t \d t\,
|\dot{f}|$. It is therefore clear that, for
sufficiently regular $f(t)$, there is a function
$C(t)$, independent of $n$, such that the absolute
value of the solutions (\ref{lin-sol}) is bounded by
$C(t)/n$. To reach this conclusion, one only needs
that the function $f(t)$ is differentiable (so that
$\dot{f}$ exists) and its derivative is integrable
in every interval $[t_0,t]\subset \mathbb{I}$ [for
instance, these conditions are satisfied if $f(t)$ is
a $C^1$ function in $\mathbb{I}$]. Then, in
particular, the sequence $\bar{W}_{n}(t)$ tends to
zero in the asymptotic limit of infinite $n$, $\forall
t\in \mathbb{I}$.

We now return to the original differential equation
(\ref{car-w}). Given the behavior of the functions
(\ref{lin-sol}), the quadratic term $\bar{W}^{2}_{n}$
is bounded in absolute value by $C(t)^2/n^2$, and is
hence negligible, in particular compared with the
linear term in that equation. Therefore, the functions
$\bar{W}_{n}(t)$ defined in formula (\ref{lin-sol})
can be taken as asymptotic solutions, in the limit of
large $n$, to Eq. (\ref{car-w}) up to subdominant
terms, terms which in any case do not affect the
square summability of the sequence $W_n(t)/n$.

Considering again the real and imaginary contributions
of the solutions $\dot\Theta_n$, and splitting also
$W_n$ in real and imaginary parts, $W_n=R_n+i I_n$,
one gets \begin{equation}\dot r_n(t)=\frac{R_n(t)}{n},
\qquad \dot s_n(t)=-1+\frac{I_n(t)}{n}.\end{equation}
According to our analysis, the real functions $R_n(t)$
and $I_n(t)$ tend to zero in the asymptotic limit of
large $n$, for all allowed values of the time $t$
$^{\footnotemark[2]}$\footnotetext[2]{Of course, a
more detailed analysis of the asymptotic behavior of
the solutions $\dot \Theta_n$ can be performed, as Eq.
(\ref{newze2}) already suggests. However, for our
purposes the current estimate is sufficient.}.

It is now a simple exercise to check that the
coefficients $\beta_n(t,t_0)$, given in Eq.
(\ref{beta}), are square summable for all $t$, since
the leading term of $|\beta_n|$ in the asymptotic
limit of large $n$ is just $(R_n+iI_n)/(2n)$. This
proves that, for a free real scalar field $\xi$ on
$S^{1}$, or an axisymmetric one on $S^2$, which is
subject to a time dependent potential
$V(\xi)=f(t)\xi^{2}/2$ (or, equivalently, with a time
dependent mass), there exists at least one Fock
representation in which the dynamics is implemented as
a unitary transformation. This Fock representation is
the one naturally associated with the massless scalar
field with vanishing potential. In particular, this
representation is $S^{1}$-invariant, i.e., it is
defined by a complex structure $J_0$ which is
invariant under the action of the group of
$S^{1}$-translations (\ref{s1-trans}) -- if we
consider the case of a field on $S^2$ instead, the
invariance is under the group $SO(3)$. This means that
the representation also provides us with a unitary
implementation of the symmetry group of the field
equations.

A natural question is whether or not the above result
holds for other nonequivalent Fock representations
which satisfy as well the requirement of invariance
under the symmetries of the field equations. An answer
in the positive would imply that one cannot pick out a
preferred $S^{1}$-invariant Fock representation [or an
$SO(3)$-invariant one for axisymmetric fields on the
sphere] by demanding the unitary implementation of the
dynamics. In contrast, if the answer is in the
negative, the $J_0$-Fock representation would be
confirmed as the unique (up to unitary equivalence)
Fock representation which is invariant under the
symmetry group of the field equations and fulfills the
demand of allowing a unitary quantum evolution. In the
next section we will show that this is indeed the
case.

\section{Uniqueness of the quantization}
\label{sec:unique}

Since, for the analyzed case of a free scalar field on
the circle with a time dependent mass or potential, we
are interested just in $S^{1}$-invariant Fock
representations, we will consider only complex
structures $J$ that are invariant under the group of
translations (\ref{s1-trans}):
$J=T_{\alpha}^{-1}JT_{\alpha}$ $\forall \alpha \in
S^{1}$. We will refer to such complex structures
simply as invariant ones. Now, given a
$\Omega$-compatible invariant complex structure $J$,
it can be shown \cite{ccmv} that it is related to
$J_0$ via $J=K_{J}J_0 K_{J}^{-1}$, where $K_J$ is a
block diagonal symplectic transformation, with
$4\times 4$ blocks of the form
\begin{eqnarray}
(K_J)_n=\left( \begin{array}{cc} ({\cal K}_J)_n & 0 \\
0 &  ({\cal K}_J)_n \end{array} \right),&& \quad
({\cal K}_J)_n= \left( \begin{array}{cc} \kappa_n &
\lambda_n
\\ \lambda_n^* & \kappa_n^* \end{array} \right)
\nonumber\\ |\kappa_n|^2-|\lambda_n|^2&=&1, \qquad
\forall n\in \mathbb{N}^+ .
\end{eqnarray}
The fact that $|\kappa_n|^2-|\lambda_n|^2=1$, implies
that $|\kappa_n|\geq 1$ and $|\kappa_n|\geq
|\lambda_n|$. Consequently, in particular, the
sequence $\{\lambda_n/ \kappa_n^*\}$ is bounded.

On the other hand, for the alternate case of an
axisymmetric field on $S^2$, the complex structures
that descend from $SO(3)$-invariant ones were
discussed in Ref. \cite{BVV2}. Besides, it was shown
in Ref.\cite{cmv3} that these invariant complex
structures can be parametrized in the same way as the
$S^1$-invariant ones. Using this common
parametrization, all of the following discussion
applies as well for this other family of field models.

Let us return to the mainstream of our argumentation.
Given a symplectic transformation $R$, it is not
difficult to see that it admits a unitary
implementation with respect to the complex structure
$J=K_{J}J_0 K_{J}^{-1}$ if and only if
$K_{J}^{-1}RK_{J}$ is unitarily implementable with
respect to $J_0$. Thus, the time evolution ${\cal{U}}$
(specified by the sequence of matrices
$\{{\cal{U}}_n\}$) will be unitarily implementable
with respect to the Fock representation determined by
$J=K_{J}J_0 K_{J}^{-1}$ if and only if the $J_0$-Fock
representation admits a unitary implementation of the
symplectic map $K_{J}^{-1}{\cal{U}} K_{J}$. This last
condition amounts to the square summability of the
sequences \be \label{beta-J}
\beta^J_n(t,t_0)=(\kappa_n^*)^2\beta_n
(t,t_0)-\lambda_n^2\beta_n^* (t,t_0)+2 i
\kappa_n^*\lambda_n {\rm{Im}}[\alpha_n(t,t_0)]\, ,
\qquad \forall t\in \mathbb{I}, \ee where $\alpha_n$
and $\beta_n$ are the Bogoliubov coefficients
corresponding to $J_0$, given in Eqs. (\ref{alpha})
and (\ref{beta}). Summarizing, a different Fock
representation, defined by a different invariant
complex structure $J=K_{J}J_0 K_{J}^{-1}$, allows a
unitary implementation of the $\xi$ field dynamics if
and only if the sequence (\ref{beta-J}) is square
summable at every instant of time $t$ in the domain
$\mathbb{I}$.

On the other hand, we recall that the Fock
representation specified by $J=K_{J}J_0 K_{J}^{-1}$
and the $J_0$-Fock representation are unitarily
equivalent if and only if the sequence $\{\lambda_n\}$
is square summable (details on this point can be found
in Ref. \cite{ccmv}). In the rest of this section, we
will demonstrate that, if the sequences
$\{\beta^J_n(t,t_0)\}$ $\forall t\in\mathbb{I}$ are
square summable, the same must necessarily happen to
the sequence $\{\lambda_n\}$. This will prove that the
$J_0$-Fock representation is in fact unique, up to
unitary equivalence.

Let us suppose that the dynamics, for some function
$f(t)$, is unitarily implemented in the invariant Fock
representation determined by the complex structure
$J$; that is, let us suppose that
$\{\beta^J_n(t,t_0)\}$ is square summable {$\forall
t\in\mathbb{I}$}. Since $|\kappa_n|>1$, it then
follows that the sequence provided by\be
\frac{\beta^J_n(t,t_0)}{(\kappa_n^*)^2}=\beta_n
(t,t_0)-\left(\frac{\lambda_n}{\kappa_n^*}\right)^{2}
\beta_n^* (t,t_0)+2 i
\left(\frac{\lambda_n}{\kappa_n^*}\right)
{\rm{Im}}[\alpha_n(t,t_0)] \ee is also square
summable. Moreover, we already know that the sequence
$\{\lambda_n/ \kappa_n^*\}$ is bounded and that the
sequence $\{\beta_n (t,t_0)\}$ is square summable
(this was shown in the previous section). These facts
guarantee then that $\{\beta_n
(t,t_0)-(\lambda_n/\kappa_n^*)^2\beta_n^* (t,t_0)\}$
is square summable. Since the space of square summable
sequences is a linear space, one is led to conclude
that the sequence $\{\left(\lambda_n/\kappa_n^*\right)
{\rm{Im}}[\alpha_n(t,t_0)]\}$ has to be square
summable as well.

The analysis that we presented in Sec. \ref{sec:unit}
to show the square summability of $\{\beta_n(t,t_0)\}$
at all instants of time $t$ can now be applied to
check that the sequence
$\{{\rm{Im}}[\alpha_n(t,t_0)]+\sin(n(t-t_0))\}$ is
also square summable. Thus, from the bound on
$\lambda_n/\kappa_n^*$ and using linearity, we
conclude that $\{(\lambda_n/\kappa_n^*)\sin(nT)\}$ is
also a square summable sequence. Here, we have called
$T=t-t_0$ in order to simplify the notation.
Therefore, the function \be g(T):=\lim_{M\to
\infty}\sum_{n=1}^{M}\frac{|\lambda_n|^{2}}
{|\kappa_n|^{2}}\sin^{2}(nT) \ee exists for all $T$ in
the interval $\mathbb{\bar{I}}$, obtained from
$\mathbb{I}$ after a negative shift by $t_0$. In
particular, the function $g(T)$ is well defined at
least on some closed subinterval of the form
$\mathbb{\bar{I}}_L=[a,a+L]\subseteq \mathbb{\bar{I}}$
(for a suitable choice of the time $a$), where $L$ is
some finite number strictly smaller than the length of
$\mathbb{I}$. Related to this number $L$, let us
introduce also, for later use, a fixed positive
integer $n_0$ such that the product $n_0L$ is larger
than the unity, a condition that can always be
fulfilled.

We can now apply Luzin's theorem \cite{luzin}, which
ensures that, for every $\delta>0$, there exist: i) a
measurable set $E_{\delta} \subset \mathbb{\bar{I}}_L$
such that its complement $\bar{E}_{\delta}$ with
respect to $\mathbb{\bar{I}}_L$ satisfies
{$\int_{\bar{E}_{\delta}}\d t<\delta$}, and ii) a
function $F_{\delta}(T)$, continuous on
$\mathbb{\bar{I}}_L$, which coincides with $g(T)$ in
$E_{\delta}$. We then get \be
\sum_{n=1}^{M}\frac{|\lambda_n|^{2}}{|\kappa_n|^{2}}
\int_{E_{\delta}}\sin^{2}(nT)\d T\leq
\int_{E_{\delta}}g(T)\d T=:I_{\delta}, \qquad \forall
M \in \mathbb{N}^+ ,\ee where
$I_{\delta}=\int_{E_{\delta}}F_{\delta}(T)\d T$ is
some finite number. Since \be
\int_{E_{\delta}}\sin^{2}(nT)\d
T=\int_{\mathbb{\bar{I}}_L} \sin^{2}(nT) \d
T-\int_{\bar{E}_{\delta}}\sin^{2}(nT) \d T\geq
\frac{L}{2}-\frac{1}{2n_0}-\delta, \qquad \forall
n\geq n_0, \ee we have that \be \label{bound}
\sum_{n=n_0}^{M}\frac{|\lambda_n|^{2}}{|\kappa_n|^{2}}
\leq \frac{2 n_0 I_{\delta}}{n_0 L-1-2 n_0 \delta}
\qquad \forall M > n_0. \ee Here, we have used that it
is possible to choose $2\delta < (L-1/n_0)$, that
\begin{equation}
\int_{\mathbb{\bar{I}}_L} \sin^{2}(nT) \d T=
\frac{L}{2}-\frac{\sin {\big[2n(a+L)\big]}}{4n}
+\frac{\sin {(2na)}}{4n}\geq \frac{L}{2}- \frac{1}{2n}
\qquad \forall n\in \mathbb{N}^+,
\end{equation}
and that $(L-1/n)\geq (L-1/n_0)$ $\forall n\geq n_0$.
Let us emphasize that Eq. (\ref{bound}) is valid for
arbitrary large $M$. Then, the right hand side of that
equation, which does not depend on $M$, provides a
bound to the increasing sequence of partial sums
$\{\sum_{n=n_0}^{M}(|\lambda_n|^{2}/|\kappa_n|^{2})\}$,
where $n_0$ is fixed. As a result, the sequence
$\{\lambda_n/\kappa_n\}$ is necessarily square
summable.

Employing the square summability of
$\{\lambda_n/\kappa_n\}$ and the identity
$|\kappa_n|^2-|\lambda_n|^2=1$, it is straightforward
to see that the sequence $\{\kappa_n\}$ is bounded.
Thus, the sequence
$\{\lambda_n=\kappa_n(\lambda_n/\kappa_n)\}$ is square
summable, as we wanted to show. Therefore, we conclude
that the $J_0$-Fock representation is the
{\sl{unique}} (up to unitary equivalence) invariant
quantum description in which the dynamics is
implemented as a unitary transformation and,
consequently, the unique invariant quantum theory
where the Schr\"{o}dinger picture can be consistently
defined.

\section{Discussion and conclusions}

It is well known that, in contrast to systems with a
finite-dimensional linear phase space, there are
inequivalent representations of the canonical
commutation relations in quantum field theory
$^{\footnotemark[3]}$\footnotetext[3]{We mean
representations of the regular type, i.e., giving rise
to irreducible representations of the Weyl relations
satisfying the standard criteria of weak continuity.}.
Clearly, this raises the issue of which choice of
representation, if any, is the adequate one for a
given classical field theory. Since, in addition to
its kinematics, a field theory is characterized by its
dynamics and its group of symmetries, it is most
natural to take into account these ingredients and let
them play a fundamental role when elucidating the
appropriate quantum representation. In fact, from a
physical point of view, it is highly questionable that
one could accept as a satisfactory quantization of the
system a representation of the canonical commutation
relations which fails to produce a unitary
implementation of the dynamics, or of the symmetries.

Therefore, given a classical field theory, the first
fundamental issue that arises regarding its
quantization is whether there exists or not a quantum
representation with a proper unitary dynamics and
symmetry group. Note that, for Poincar\'e invariant
theories, dynamics and symmetries are unified under
the Poincar\'e group, and one looks in fact for
unitary implementations precisely of that group. In
the particular case of free scalar fields on Minkowski
spacetime, representations with the desired properties
are known to exist, of course. These are the familiar
free field representations, defined by Poincar\'e
invariant complex structures, and are distinguished by
the value of the mass. This means that, for each mass
$m$, there exists a different, unitarily inequivalent,
representation of the canonical commutation relations.
The dynamics of the massive field with mass equal to
$m$ is unitarily implemented on the corresponding
$m$-representation, but not on any of the distinct
representations defined by $m'\not = m$.

The situation described above is a neat example of the
necessity of invoking the dynamics, and the
symmetries, in the quantization process. It
illustrates the general belief that, in quantum field
theory, the representation depends on the dynamics.
One can, furthermore, expect that dynamical (or energy
related) considerations might fix the representation
uniquely $^{\footnotemark[4]}$\footnotetext[4]{Leaving
aside situations of spontaneous symmetry breaking.}
(up to unitary equivalence). This is indeed the case
for free fields.

In the present work, we have carried out the analysis
of the canonical quantization of a free scalar field
$\xi(t,\theta)$ on the circle in the presence of an
explicitly time dependent potential
$V(\xi)=f(t)\xi^{2}/2$, interpretable as a time
dependent mass. With respect to the set of field
theories commented above, there are two major
differences. On the one hand, the effective space
where the field lives is now the compact space $S^1$,
which is an important simplification. On the other
hand, the field is not truly free, as the ``mass''
term depends on time. Time translation invariance is
therefore lost and, though linear, the dynamics is
non-trivial. Thus, the existence of a representation
with unitary dynamics is not granted {\em a priori}.

The first result that we have demonstrated is that,
for the considered type of field theories, there
exists indeed a representation which allows a unitary
implementation of the dynamics, namely, the
representation which is naturally associated with the
massless free field on $S^1$. Moreover, this result
holds for all sufficiently regular functions $f(t)$
[it suffices that $f(t)$ is differentiable and its
derivative is integrable in every compact subinterval
of the domain of definition].

This result is better understood by reminding the
reader that the above mentioned inequivalence between
the free field representations in Minkowski spacetime
is due to the long range behavior (see Ref.
\cite{mtv}), which is absent in the $S^1$ case.
Actually, the representations of the canonical
commutation relations associated with free fields on
$S^1$ {\em are} all unitarily equivalent \cite{ccmv},
for any value of the mass. We have shown that,
remarkably, the zero mass representation (and
therefore the free field representation for any other
value of the mass) also supports the dynamics of our
field for every choice of the (regular) function
$f(t)$. In this sense, the free field representation
emerges in the compact case as a fixed stage where (at
least some) different representations of interest are
simultaneously realized. Moreover, the free field
representation is defined by a complex structure which
is invariant under $S^1$-translations, and therefore
carries as well a unitary implementation of that
symmetry group.

The second result that we  have proved is the
uniqueness of the quantum representation. In addition
to a unitary implementation of the dynamics, we
require that the representation is defined by a
$S^1$-invariant complex structure. Under these
conditions, we have shown that any representation
which supports a unitary dynamics for the field $\xi$,
for a given function $f(t)$, is unitarily equivalent
to the massless free field representation. Thus, our
conditions provide a successful uniqueness criterion.

It may be worth commenting on a couple of points to
clarify the significance of this result. On the one
hand, the set of representations defined by
$S^1$-invariant complex structures is quite large and
definitely contains different unitary equivalence
classes. Therefore, the fact that unitary
implementation of the dynamics selects precisely one
equivalence class has to be considered non-trivial. Of
course, it would be desirable to extend our results in
order to include representations defined by general
$S^1$-invariant algebraic states, rather than just
states induced by complex structures. However, since
our symmetry group of (constant) $S^1$-translations is
relatively small, that would leave us with a huge set
of states with, at least to our knowledge, no
manageable characterization. On the other hand, we
note that the condition of $S^1$-invariance, though
not strictly necessary for the unitary implementation
of $S^1$-translations, is certainly natural, and
follows the general procedure for the implementation
of symmetries (provided that invariant states exist,
of course).

Although we have centered our attention on the $S^1$
case, mainly for simplicity, we have seen that all our
results extend to the case of axisymmetric fields on
the two-sphere, with the proper adaptations [for
instance, the symmetry group would be $SO(3)$ rather
than $S^{1}$]. In this regard, let us stress that the
general analysis here presented is motivated by the
current interest in the quantization of symmetry
reductions of General Relativity in the presence of
two commuting spacelike Killing vectors. Actually, the
two considered cases of the field on the circle and
the axisymmetric field on the sphere cover all
symmetry reductions of this kind found in cosmology,
when one restricts to compact spatial sections,
provided that the Killing vectors are hypersurface
orthogonal (linear polarization). Such symmetry
reductions correspond to the families of Gowdy
spacetimes. Thus, our analysis includes the recent
treatments of the quantum Gowdy cosmologies:
specifically, the linearly polarized $T^{3}$ model
\cite{CM,ccm1,ccm2,ccmv,cmv} is described by a scalar
field $\xi$ on $S^{1}$ with ${f(t)}=1/(2t)^2$, whereas
the Gowdy $S^{1}\times S^{2}$ and $S^{3}$ models
\cite{BVV1,BVV2,cmv3} admit a description in terms of
an axisymmetric field $\xi$ on $S^{2}$ with
$f(t)=(1+\csc^{2}t)/4$. With the present unified
treatment, we expect to have contributed both to a
better understanding of the previously obtained
results, and to an extension of them which can find
applications in other symmetry reductions of General
Relativity or, more generally, in quantum field theory
on curved backgrounds.


\section*{Acknowledgements}
\noindent J.C. would like to thank R.P. Mart\'{\i}nez
y Romero for comments and conversations. This work was
supported by the Spanish MICINN Project
FIS2008-06078-C03-03 and Consolider-Ingenio 2010
Program CPAN (CSD2007-00042) and the DGAPA-UNAM
(Mexico) grant No. IN108309-3.



\end{document}